\documentclass[10pt]{article} 

\usepackage{pdfpages}
\usepackage{epsfig}
\usepackage{graphicx}
\usepackage{amsmath, amsthm, amssymb}
\usepackage{setspace}  
\usepackage{verbatim} 

\newcommand{\be}{\begin{equation}}
\newcommand{\ee}{\end{equation}} 
\newcommand{\lb}{\label}
\newcommand{\OL}{\overline}

\newcommand{\ba}{{\bf a}}

\newcommand{\bk}{{\bf k}}

\newcommand{\br}{{\bf r}}
\newcommand{\bu}{{\bf u}}

\newcommand{\bx}{{\bf x}}

\newcommand{\bz}{{\bf z}}

\newcommand{\bJ}{{\bf J}}

\newcommand{\bomega}{{\mbox{\boldmath $\omega$}}}

\newcommand{\grad}{{\mbox{\boldmath $\nabla$}}}
\newcommand{\bdot}{{\mbox{\boldmath $\cdot$}}}

\newcommand{\btimes}{{\mbox{\boldmath $\times$}}}

\begin{document}

\baselineskip=18pt
\begin{center}
\begin{LARGE}
{\bf Joint downscale fluxes of energy and potential enstrophy in rotating stratified Boussinesq flows}\\
\end{LARGE}

\bigskip
\bigskip

Hussein Aluie$^{1,2}$ and Susan Kurien$^{2,3}$\\
{\it
  $^{1}$ Center for Nonlinear Studies\\
$^{2}$ Applied Mathematics and Plasma Physics (T-5)\\
  Theoretical Division, Los Alamos National Laboratory, Los Alamos, New Mexico 87545, USA\\
$^{3}$ New Mexico Consortium, Los Alamos, New Mexico 87544, USA}

\bigskip
\bigskip

\begin{abstract}
We employ a coarse-graining approach to analyze nonlinear
  cascades in Boussinesq flows using high-resolution simulation
  data. We derive budgets which resolve the evolution of energy and
  potential enstrophy simultaneously in space and in scale.  We then
  use numerical simulations of Boussinesq flows, with forcing in the
  large-scales, and fixed rotation and stable stratification along the
  vertical axis, to study the inter-scale flux of energy and potential
  enstrophy in three different regimes of stratification and rotation:
  (i) strong rotation and moderate stratification, (ii) moderate
  rotation and strong stratification, and (iii) equally strong
  stratification and rotation.  In all three cases, we observe
  constant fluxes of both global invariants, the mean energy and mean
  potential enstrophy, from large to small scales. The existence of
  constant potential enstrophy flux ranges provides the first direct
  empirical evidence in support of the notion of a cascade of
  potential enstrophy.  The persistent forward cascade of the two
  invariants reflects a marked departure of these flows from
  two-dimensional turbulence.
  \end{abstract}
\end{center}

\vspace{0.5cm}

~~~{\bf Key Words:} Turbulent flows; Stratified flows; Rotating flows.
\clearpage

\section{Introduction}
The Kolmogorov \cite{Kolmogorov41a} and Kraichnan \cite{Kraichnan67}
theories of three- and two-dimensional Navier-Stokes turbulence have
served as a benchmark in the understanding of fluid turbulence and as
fundamental tests for the accuracy of simulations. The Boussinesq
approximation of the compressible Navier-Stokes equations in a
rotating frame give a fairly accurate description of the flow dynamics
over much of the Earth's oceans and atmosphere but are prohibitively
expensive to simulate in detail over global scales. Guided by the
success of the Kolmogorov/Kraichnan theories, it would be useful to
develop a statistical phenomenology of the small scales of Boussinesq
flows to gain an understanding of the physics, serve as benchmarks to
test simulations, and offer parameterizations which could eventually
be useful in practical modeling of geophysical flows. In addition to
global energy, the inviscid Boussinesq equations conserve local
potential vorticity (hereafter, PV) and global potential enstrophy, thus offering more
complexity than incompressible Navier-Stokes dynamics.
Charney \cite{Charney71} addressed one limiting case of the Boussinesq
approximation, namely the quasi-geostrophic limit of strong rotation
and strong stratification, and showed that the conservation of both
energy and the quadratic potential enstrophy in such flows constrained
energy to cascade to the large scales as in 2D turbulence
\cite{Kraichnan67}.

Conservation of potential enstrophy is believed to play a fundamental
role in the dynamics of the atmosphere and oceans
\cite{Vallis06}. Understanding its function in nonlinear scale
interactions would appear to be essential for extending Kolmogorov's
theory to Boussinesq flows with rotation and stratification. Herring
et al. \cite{Herringetal94} studied the cascade properties of
potential enstrophy in turbulence simulations with a passive scalar in
the absence of rotation and stratification. 
The authors concluded that because
potential enstrophy in their simulation is not a quadratic, but 
a quartic invariant, the
usual Kolmogorov-like arguments for a cascade and an inertial range do
not apply. In fact, their potential enstrophy budget shows the direct
action of viscous-diffusion terms at \emph{all scales}, even in the
limit of very small viscosity and diffusivity.
Their figure 15 shows that potential enstrophy dissipation 
peaks at the largest scales (see also their figure 14 and the discussion 
on pp. 37 and 43). Thus, a pure inertial range of potential enstrophy flux 
is precluded because of contamination by dissipation at all scales.
However for Boussinesq flows, with strong
rotation and/or strong stratification, Kurien et
al. \cite{Kurienetal06} (hereafter, KSW06) derived
analytically, starting from the evolution of the
  two-point correlation of potential vorticity, a flux law for
potential enstrophy which is analogous to Kolmogorov's 4/5-law for
energy flux in 3D incompressible turbulence.
The so-called 2/3-law of KSW06 implies that an inertial cascade of
potential enstrophy can exist in three limiting cases: (i) strong
rotation with moderate stratification (ii) moderate rotation with
strong stratification and (iii) strong rotation and strong
stratification. By ``moderate'' we mean that the rotation
  (stratification) frequency is of the same order as the non-linear
  turbulence frequency; whereas by ``strong'' we mean that the
  respective frequencies are much faster than the non-linear
  timescale. In  the above three regimes,
KSW06 showed that potential enstrophy, generally a quartic quantity,
becomes quadratic which results in
the localization of viscous-diffusion terms to the smallest
scales, thus allowing for an inertial range of scales
  dominated by the flux of potential enstrophy. Furthermore, KSW06
suggested that in the absence of strong rotation and/or strong
stratification, when potential enstrophy reverts to
  being quartic, viscous-diffusion effects may contaminate all
scales, in agreement with the conclusions of \cite{Herringetal94}.

The existence of an inertial cascade range for potential enstrophy is
far from obvious and remains an unsettled issue. To date, there has
been no empirical demonstration of KSW06's results on the constant
flux range of potential enstrophy in Boussinesq flows. In
\cite{Kurienetal08} a phenomenology and supporting data from
Boussinesq simulations with equally strong (non-dimensional) rotation
and stratification were presented to show that conservation of
quadratic potential enstrophy could constrain the spectral
distribution of energy in the large wavenumbers. In \cite{Kurien11}
the analysis and phenomenology of \cite{Kurienetal08} was extended to
include the two other limiting cases (i) and (ii) above, to show that
indeed in all parameter regimes with linear PV, the downscale flux of
(quadratic) potential enstrophy can constrain the scale-distribution
of energy. While the two studies \cite{Kurienetal08,Kurien11} 
did not measure potential enstrophy flux, their numerical findings confirmed 
phenomenologically predicted scalings of the energy
spectra assuming constant downscale fluxes of potential
enstrophy. Motivated partially by those results, the present work uses 
data from \cite{Kurien11} to directly measure the cascade of potential enstrophy
for the first time.

Some previous studies have highlighted the
  interest in energy fluxes. The results from
  \cite{Lindborg06,Brethouweretal07} of purely stratified flow
  computed in small-aspect-ratio domains with forcing of the
  horizontal velocity only, did not show convincing constant energy
  fluxes (see figure 18 of \cite{Lindborg06} and figure 15 of \cite{Brethouweretal07} 
  which plot energy flux on a log-log scale). On the other hand \cite{Waite11} provides 
  evidence of scale-independent fluxes of kinetic and potential energy in purely
  stratified flow in unit aspect-ratio. It should be noted that these studies 
used forcing, stratification, and domain aspect-ratio different from ours.
Furthermore, the simulations of \cite{Lindborg06,Brethouweretal07,Waite11} 
were not reported to be in the linear PV regime and, therefore, our results
  may not apply to their flows.
Most importantly, none of these previous studies computed or analyzed the potential enstrophy flux, 
which constitutes an essential part of our work.

In this Letter, we present a very general framework for analyzing
nonlinear scale interactions in Boussinesq flows. The coarse-graining
approach we utilize allows for probing the dynamics simultaneously in
space and in scale.  Motivated by the work of
\cite{Kurienetal06,Kurienetal08,Kurien11}, we then measure fluxes of
energy and potential enstrophy across scales from simulations in three
distinct limits of rotation and stratification. Our results show
constant and positive fluxes of the two quadratic invariants,
indicating simultaneous persistent downscale cascades of both
quantities in all three cases.  Our measurements of potential
enstrophy flux are a novel contribution of this Letter and constitute
the first empirical confirmation of analytical results by
\cite{Kurienetal06}. Furthermore, our evidence of a scale-independent
energy flux is significant because it conveys that a cascade should
persist to arbitrarily small scales at asymptotically high simulation
resolutions.

\section{Boussinesq dynamics}
We study stably stratified Boussinesq flows
in a rotating frame. The dynamics is described by momentum (\ref{momentum}) and 
active scalar (\ref{scalar}) equations:
\begin{eqnarray}
\hspace{-1.5mm}\partial_t \bu + (\bu\bdot\grad)\bu \hspace{-3mm}&=&\hspace{-3mm}-\grad p  -f \hat{\bz}\btimes\bu - N \theta\hat{\bz} + \nu\nabla^2\bu 
+ {\boldsymbol {\mathcal F}}^u,
\lb{momentum} \\
\hspace{-1.5mm}\partial_t \theta + (\bu\bdot\grad)\theta \hspace{-3mm}&=&\hspace{-3mm}  N u_z + \kappa \nabla ^2\theta
+ {\mathcal F}^\theta.
\lb{scalar}
\end{eqnarray}
Here, $\bu$ is a solenoidal velocity field, $\grad\bdot\bu = 0$, whose vertical component is $u_z$.
The effective pressure is $p$, and ${\boldsymbol {\mathcal F}}^u$,${\mathcal F}^\theta$
are external forces. Gravity, $g$,  is constant and in the $-\hat{\bz}$ direction.
Total density is given by $\rho_T(\bx) = \rho_0 -bz +\rho(\bx)$, such that
$|\rho(\bx)| \ll |bz|$ and $|\rho(\bx)| \ll \rho_0$, where $\rho_0$ is a constant background density, $b$ is constant 
and positive for stable stratification, and $\rho(\bx)$ is the fluctuating density field with zero mean.
The normalized density, $\theta(\bx) = \sqrt{g/b\rho_0} \rho(\bx)$, has units of velocity. 
For a constant rotation rate $\Omega$ about the z-axis, the Coriolis parameter is $f=2\Omega$.
The Brunt-V\"ais\"al\"a frequency is $N=\sqrt{gb/\rho_0}$, kinematic viscosity is $\nu=\mu/\rho_0$,
and mass diffusivity is $\kappa$. In this paper, we only study flows with Prandtl number 
$Pr = \nu/\kappa = {\cal O}(1)$. Relevant non-dimensional parameters are Rossby number,
$Ro = f_{nl}/f$, and Froude number, $Fr = f_{nl}/N$, where we define the characteristic non-linear 
frequency as $f_{nl} = (\epsilon_f k_f^2)^{1/3}$, for a given energy injection rate $\epsilon_f$
at wavenumber $k_f$ (see \cite{SmithWaleffe02,Kurienetal06,Kurien11}).

The dynamics of inviscid and unforced Boussinesq flows 
(such that $\nu=\kappa={\mathcal F}^\theta = {\mathcal F}^u_i = 0$) is constrained by the
conservation of potential vorticity, $q(\bx) = \frac{N}{b}\bomega_a\bdot\grad\rho_T$,
following material flow particles,
$ D_t q = \partial_t q + (\bu\bdot\grad)q = 0$.
Here, absolute vorticity is $\bomega_a = \bomega + f\hat{\bz}$ and local vorticity is 
$\bomega = \grad\btimes\bu$. PV may be written in terms of $\bomega$ and $\theta$
as \be q(\bx) = f\partial_z\theta - N\bomega\bdot\hat{\bz} + \bomega\bdot\grad\theta -fN  .
\lb{PVexpand}\ee
The first two terms are linear and dominate over the quadratic term, $\bomega\bdot\grad\theta$,
in the limit of large $f$ and/or large $N$. The constant part in (\ref{PVexpand}) does not 
participate in the dynamics and can, therefore, be neglected \cite{SmithWaleffe02}.

In addition to conservation of PV, the flow is constrained by the global conservation of 
potential enstrophy, $Q = \frac{1}{2} q^2$, such that
$\frac{d}{dt} \langle Q\rangle = 0,$
where $\langle\dots\rangle = \frac{1}{V}\int_V d^3\bx(\dots)$ is a space average. Another quadratic
 invariant of the inviscid dynamics is total mean energy, $E_T =  \frac{1}{2} \langle|\bu|^2 + |\theta|^2\rangle$,
such that $ \frac{d}{dt} E_T = 0.$

\section{Numerical data}
The Sandia-LANL DNS code was used to perform pseudo-spectral calculations of the Boussinesq equations
(\ref{momentum})-(\ref{scalar}) on grids of $640^3$ points in unit
aspect-ratio domains. 
The time-stepping is 4th-order Runge-Kutta and the fastest linear wave
frequencies are resolved with at least five timesteps per wave
period. The diffusion of both momentum and density (scalar) is modeled
by hyperviscosity of laplacian to the 8th-power.  The
coefficient of the hyperviscous diffusion term
is chosen dynamically such that the energy
in the largest wavenumber shell (smallest resolved scale) is dissipated
at each time-step \cite{Chasnov94,SmithWaleffe02}. This choice ensures
that one does not have to guess the coefficient {\it a priori} and is a
way to allow the flow itself to determine the magnitude of the
diffusion. Hyperviscosity is a standard dissipation model that has long been used in studies of rotating and/or stratified flows \cite{Lindborg06,Brethouweretal07,Waite11,SmithWaleffe02}. In principle, hyperviscosity can lead to thermalization and isotropy at the smallest scales \cite{Frischetal08}, however our results on the inertial-range cascades are robust and unaffected by the small-scale dissipation model \cite{AluieKurien_long}.
Stochastic forcing is incompressible and equipartitioned between
the three velocity components and $\theta$. The forcing spectrum is
peaked at $k_f = 4\pm 1$, for large scale forcing. We use the two-thirds dealiasing rule.
These data were reported in \cite{Kurienetal08,Kurien11},
where further computational details may be found.

\begin{table}[h]
\centering
\begin{tabular}{|cccc|}
  \hline
  Run  & resolution  & $f~(Ro)$ & $N~(Fr)$\\ 
  \hline
  Rs & $640^3$ & $3000~(.002)$ & $14~(.4)$ \\
  rS & $640^3$ & $14~(.4)$ & $3000~(.002)$  \\
  RS & $640^3$ &$3000~(.002)$ & $3000~(.002)$  \\
  \hline
\end{tabular}
\caption{Parameters of the Boussinesq simulation data.}
\label{param_table}
\end{table}
We analyze three sets of simulations corresponding to three extreme flow regimes summarized in
Table \ref{param_table}.
The first, Rs, is a flow under strong rotation and moderate stratification, $f/N \gg 1$.
The second, rS, is a flow under moderate rotation but strong stratification, $f/N \ll 1$.
The third, RS, is a flow under strong rotation and strong stratification such that $f=N$.
Figure \ref{Fig:PV2timeseries} shows that in all three cases, $\langle Q\rangle$ is well approximated by (one half)
the square of the corresponding linear PV to within $3\%$ or better (see \cite{Kurienetal06,Kurien11}).
Figure \ref{Fig:PV2timeseries} is evidence that
our simulations are in regimes of strong rotation and/or strong stratification.
We analyze snapshots of the flow at late times when $\langle Q\rangle$ along with
small-scale energy spectra (at wavenumbers $k\ge 6$) have reached a statistically steady state.
The total energy, however, continues to grow due to an accumulation at the largest scales.

\begin{figure}[htpb]
\centering
\includegraphics[totalheight=0.35\textheight,width=0.6\textwidth]{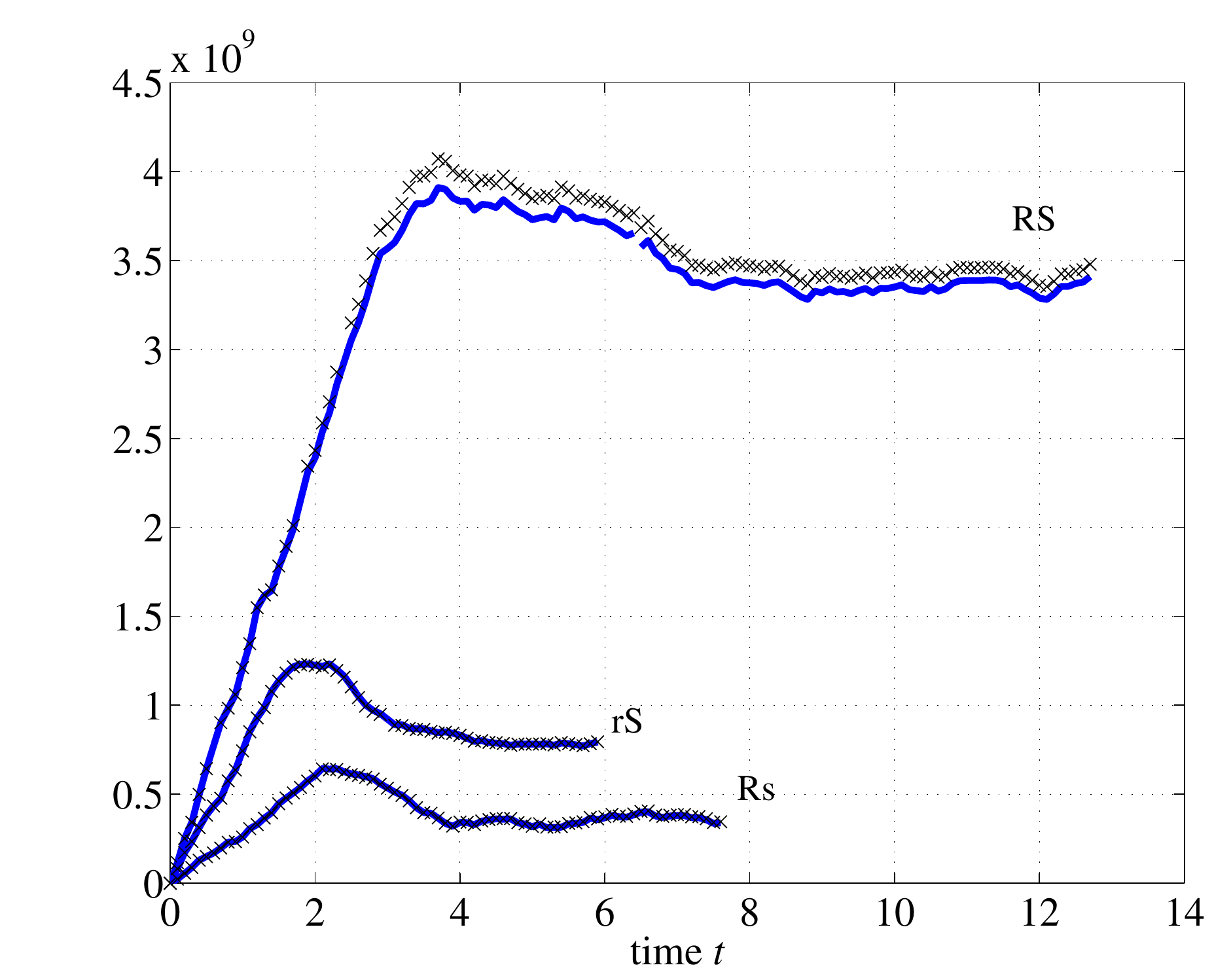}
\caption{Time-series of mean potential enstrophy $\langle Q\rangle$
  (solid line), and its quadratic part $\langle \widetilde Q\rangle$
  (crosses).  In run RS $\langle \widetilde Q\rangle
  =\big\langle\left[ f\partial_z\theta - N\bomega\bdot\hat{\bz}
  \right]^2\big\rangle/2$; in run rS $\langle \widetilde Q \rangle
  =\big\langle\left[ N\bomega\bdot\hat{\bz} \right]^2\big\rangle/2$;
  and in run Rs $\langle \widetilde Q \rangle =\big\langle\left[
    f\partial_z\theta \right]^2\big\rangle/2$.  The plots
  show that $\langle Q\rangle$ reaches steady-state and that $\langle Q \rangle \simeq \langle \widetilde Q \rangle$ in all three regimes
  considered.}
\lb{Fig:PV2timeseries}\end{figure}

\section{Analyzing the cascades by coarse-graining}
Following \cite{Leonard74,Germano92,Eyink95a,Eyink05}, we use a simple filtering technique
common in the Large Eddy Simulation (LES) literature to resolve turbulent fields simultaneously 
in scale and in space. Other decompositions, such as wavelet analysis, also allow
for the simultaneous space-scale resolution and may be used to analyze nonlinear scale interactions as well.

We define a coarse-grained or (low-pass) filtered field in $d$-dimensions as  
\be
\OL \ba_\ell(\bx) = \int d^d\br~ G_\ell(\br) \ba(\bx+\br),
\lb{filtering}\ee
where $G(\br)$ is a normalized convolution kernel, $\int d^d\br ~G(\br)=1$. 
An example of such a kernel is the Gaussian function, $G(r) = \frac{1}{\sqrt{2\pi}}e^{-r^2/2}$.
Its dilation $G_\ell(\br)\equiv \ell^{-d} G(\br/\ell)$ in $d$-dimensions
has its main support in a ball of radius $\ell$. Operation (\ref{filtering}) may 
be interpreted as a local space average.
In the rest of our Letter, we shall omit subscript  $\ell$ whenever there is no ambiguity.

Applying the filtering operation (\ref{filtering}) to the dynamics (\ref{momentum})-(\ref{scalar})
yields coarse-grained equations that describe the evolution of 
$\OL{\bu}_\ell(\bx)$ and $\OL\theta_\ell(\bx)$ at every point $\bx$ in space 
and at any instant of time:
\begin{eqnarray} 
\partial_t \OL\bu + (\OL\bu\bdot\grad)\OL\bu \nonumber
&=& -\grad \OL{p}-f \hat{\bz}\btimes\OL\bu - N \OL\theta\hat{\bz} \\
& &-\grad\bdot\,\OL\tau(\bu,\bu) + \nu\nabla^2\OL\bu + \OL{{\boldsymbol {\mathcal F}}^u},
\lb{largemomentum} \\
\partial_t \OL\theta + (\OL\bu\bdot\grad)\OL\theta &=&\nonumber 
  N\, \OL{u}_z -\grad\bdot\,\OL\tau(\bu,\theta)\\
& &+ \kappa \nabla ^2\OL\theta + \OL{{\mathcal F}^\theta},
\lb{largescalar}
\end{eqnarray}
where \emph{subgrid stresses}, $\OL\tau_\ell(f,g)\equiv \OL{fg}_\ell - \OL{f}_\ell~ \OL{g}_\ell$,
are ``generalized 2nd-order moments'' \cite{Germano92} accounting 
for the influence of eliminated fluctuations at scales $<\ell$. 

The coarse-grained equations describe flow at scales $>\ell$, for arbitrary
$\ell$. The approach, therefore, allows for the simultaneous 
resolution of dynamics \emph{both in scale and in space}.
Furthermore, the approach admits intuitive physical interpretation of various terms in 
the coarse-grained balance (\ref{largemomentum})-(\ref{largescalar}) which
resemble the original governing eqs. (\ref{momentum})-(\ref{scalar}),
except for additional \emph{subgrid} terms 
which quantify the nonlinear coupling between resolved and filtered scales.
These subgrid terms depend inherently on the unresolved dynamics which has been 
filtered out. Traditional modeling efforts, such as in LES 
(see for example \cite{MeneveauKatz00}), focus on devising closures for such
terms which are plausible but whose regimes of applicability and validity are inevitably unknown.
A key feature of the formalism employed here that distinguishes it from those modeling efforts is
that it allows us to estimate the contribution of subgrid terms
as a function of the resolution scale $\ell$ through exact mathematical analysis
and direct numerical simulations (see for example \cite{EyinkAluie09,AluieEyink09,AluieEyink10,Aluie11c}). Our approach thus quantifies the coupling that exists between different scales and may be used to extract certain scale-invariant features in the dynamics.

\subsection{Large-scale energy budget}
From eqs. (\ref{largemomentum}) and (\ref{largescalar}),
it is straightforward to derive an energy budget for the large-scales, which reads
\begin{eqnarray} 
\partial_t \left(\frac{|\OL\bu|^2}{2}+\frac{|\OL\theta|^2}{2}\right) + \grad\bdot\bJ_\ell 
\hspace{-2mm}&=& \hspace{-2mm}-\Pi_\ell  -\nu|\grad\OL\bu|^2-\kappa|\grad\OL\theta|^2 
\hspace{+6mm} \lb{largeE}\\\nonumber
&& +\OL\bu\bdot\OL{{\boldsymbol {\mathcal F}}^u}  +\OL\theta~\OL{{\mathcal F}^\theta}.
\end{eqnarray}
Here, $\nu|\grad\OL\bu|^2+\kappa|\grad\OL\theta|^2$ is molecular dissipation acting
on scales $>\ell$, and $\OL\bu\bdot\OL{{\boldsymbol {\mathcal F}}^u}+\OL\theta~\OL{{\mathcal F}^\theta}$ 
is energy injected due to external stirring. 
The term $\bJ_\ell(\bx)$ represents space transport of large-scale energy
whose complete expression is deferred to an Appendix below.
Subgrid scale (SGS) flux, $\Pi_\ell(\bx)$,
accounts for the nonlinear transfer of energy from scales $>\ell$ to smaller scales: 
\be
\Pi_\ell(\bx) =-\partial_j\OL{u}_i  \OL\tau(u_i,u_j)-\partial_j\OL{\theta}  ~\OL\tau(\theta,u_j).
\lb{SGSfluxes}\ee
The SGS flux in (\ref{SGSfluxes}) is work done by large-scale velocity and scalar
gradients, $\grad\OL\bu(\bx)$ and $\grad\OL\theta(\bx)$, against subgrid stresses. 
It acts as a sink in the large-scale budget (\ref{largeE}) and accounts for the energy 
transferred across scale $\ell$ at any point $\bx$ in the flow. 
Furthermore, $\Pi_\ell(\bx)$ is Galilean invariant. Other definitions
of a flux are possible, such as $\Pi_\ell(\bx) = \OL{u}_i u_j
\partial_j(u_i-\OL{u}_i)$ (Eq. (2.52) in Frisch \cite{Frisch}), which
differs from our definition (\ref{SGSfluxes}) by a total gradient
(disregarding the scalar part). However, these alternate definitions
are not pointwise Galilean invariant, so the amount of energy
cascading at any point $\bx$ in the fluid according to such definitions would depend on the observer's velocity. 

Another physical requirement on the SGS flux $\Pi_\ell(\bx)$  is that it should vanish in the absence of fluctuations at scales smaller than $\ell$ \cite{EyinkAluie09,AluieEyink09}. For example, when $\ell= K^{-1}_{max}$, where $K_{max}$ is the maximum wavenumber in a pseudospectral simulation, there should be no cascade across $\ell$ simply because fluctuations at wavevectors $|\bk|>K_{max}$ have zero amplitude. This is satisfied by our definition (\ref{SGSfluxes}) identically at every point $\bx$ in the flow. Alternate definitions such as $\Pi_\ell(\bx) = \OL{u}_i u_j \partial_ju_i$ (eq. (6) in \cite{Mininnietal08}) fail this pointwise requirement of a flux.

\subsection{Large-scale $Q$ budget}
Similar to the momentum, scalar, and energy equations (\ref{largemomentum})-(\ref{largeE}), 
we can write down large-scale balances 
for PV and $Q$. The ``bare'' PV equation with diffusion and external forcing may 
be derived from eqs. (\ref{momentum}) and (\ref{scalar}) as
\begin{eqnarray}
\partial_t q + (\bu\bdot\grad)q \hspace{-3mm}&=& \hspace{-3mm} \lb{PV}
\nu(\grad\theta -N\hat{\bz}) \,\bdot \nabla^2\bomega_a  + \kappa \bomega_a\bdot \nabla^2\grad \theta~~\\\nonumber
&&+ (\grad\theta -N\hat{\bz})\,\bdot{{\boldsymbol {\mathcal F}}^\omega} + \bomega_a\bdot\grad{{\mathcal F}^\theta},
\end{eqnarray}
where ${{\boldsymbol {\mathcal F}}^\omega} = \grad\btimes{{\boldsymbol {\mathcal F}}^u}$. 
Applying the filtering operation to (\ref{PV}) in the limit of strong rotation and/or stratification 
(limit of large $f$ and/or $N$ such that PV is linear, $q = f\partial_z\theta -N\omega_z$), 
yields the following balance for large-scale PV 
\begin{eqnarray} \partial_t \OL{q} + (\OL{\bu}\bdot\grad)\OL{q} 
\hspace{-2mm}&=&\hspace{-2mm} -\grad\bdot\OL\tau(q,\bu)
\lb{largePV}\nonumber
-\nu \, N \, \nabla^2 \OL{\omega}_z  + \kappa \, f \, \nabla^2 \partial_z \OL{\theta}\hspace{+2mm}\\
&&- N \, \OL{{\mathcal F}^\omega}_z +  f \, \partial_z \OL{{\mathcal F}^\theta}.
\end{eqnarray}

Using (\ref{largePV}), we can now write the large-scale potential enstrophy budget in the 
limit of linear PV,
\begin{eqnarray}
\partial_t \frac{|\OL{q}|^2}{2} + \grad\bdot\bJ^{Q}
= -\Pi^{Q}  -D^{Q} + \epsilon_{inj}^{Q},
\lb{largePV2}
\end{eqnarray}
where $\bJ^{Q}(\bx)$ is space transport, 
$D^{Q}_\ell(\bx)$ is dissipation due to viscosity and diffusivity acting
directly on scales $>\ell$, and $\epsilon_{inj}^{Q}$ is the potential enstrophy injected due
to external forcing. These terms are defined in \cite{AluieKurien_long}.
$\Pi^{Q}_\ell(\bx)$ is SGS flux of potential enstrophy, 
\be
\Pi^{Q}_\ell(\bx) = -\partial_j \OL{q} ~ \OL{\tau}(q,u_j).
\lb{PV2flux}\ee
It is straightforward to verify that $\Pi^{Q}_\ell(\bx)$ is Galilean invariant
and vanishes in the absence of subgrid fluctuations.

\section{Calculating fluxes using sharp spectral filter}
We choose the so-called ``sharp spectral filter'' as our coarse-graining
kernel in the definition of fluxes.
We denote a field in a periodic domain $[0,1)^3$, coarse-grained with the 
spherically symmetric sharp-spectral filter to retain only Fourier modes $|\bk|<K$, by 
\be\ba^{<K}(\bx) \equiv \sum_{|\bk|\le K}d\bk ~\hat{\ba}(\bk) e^{i 2\pi\bk\cdot\bx}.
\lb{iso-sharp-spectral-def}\ee
This is similar to $\OL\ba_\ell(\bx)$ with $\ell\sim K^{-1}$. We omit the factor $2\pi$ in 
reference to wavenumber in this Letter. While an isotropic filter such as
(\ref{iso-sharp-spectral-def}) cannot distinguish between different directions,  
it does not average out anisotropy if present in a flow as will be clearly demonstrated in \cite{AluieKurien_long}.

Using this filter, we can discern the amount of energy and potential enstrophy 
cascading across a certain wavenumber $K$. 
For example, to analyze the energy cascade, we can compute
\be \Pi_{K}(\bx) = -\partial_j u_i^{<K} \tau^{<K}(u_i,u_j) - \partial_j \theta^{<K} \tau^{<K}(\theta,u_j) 
\lb{eFlux_spectral}\ee
as a function of $K$. 
Here, $\tau^{<K}(f,g) = (fg)^{<K} - f^{<K}g^{<K}$. 
We can also analyze potential enstrophy cascade using 
\begin{eqnarray}\Pi^{Q}_{K}(\bx) 
= -\partial_j q^{<K} \tau^{<K}(q,u_j).
\lb{qFlux_spectral}\end{eqnarray}
SGS energy flux (\ref{eFlux_spectral}) coincides with that used in
\cite{Lindborg06,Brethouweretal07}
only after space averaging. Yet, our quantity has the correct pointwise
physical properties discussed above and, therefore, allows for  
studying spatial properties of the cascades.

\section{Numerical Results}
In this Letter, we restrict our numerical investigation to spatially averaged fluxes 
using the sharp spectral filter. Figure \ref{Fig:Flux_All_iso}
shows that there is a positive and constant flux 
(y-axis shown on a linear scale to highlight true constancy) 
of total energy to small scales in all three cases of 
rotation and stratification\footnote{
The flux in the rS case is noisy at small scales 
because of a highly anisotropic cascade across wavevectors with large vertical 
component $k_z$ while being entirely suppressed across modes with large 
horizontal component $k_h$ \cite{AluieKurien_long}.
The number of such anisotropic wavevectors does not increase with $K$ and, therefore, 
does not provide the additional averaging needed to smooth out small-scale noise.}.
While all three cases show a clear downscale energy cascade,
the RS case also has a negative flux over $K\le 4$ possibly indicative
of the expected inverse cascade present in such regime 
(e.g. \cite{Bartello95,Babinetal97,EmbidMajda98}). We do not have the 
required scale-range in our simulations to say anything more definitive about the dynamics at 
scales larger than that of forcing.

We wish to emphasize the constancy of fluxes.
A constant flux indicates a \emph{persistent} non-linear transfer of energy to smaller scales, i.e.
the flow is able to sustain a cascade to \emph{arbitrarily small scales}, regardless of
how small the viscous-diffusion parameters are. The term ``cascade'' necessitates a 
flux constant in wavenumber. It is certainly possible for non-linear interactions
to yield a transient transfer to smaller scales but one which does not persist (decays to zero) 
and cannot carry the energy all the way to molecular scales. This is sometimes observed, for example,
in 2D turbulence simulations and experiments (e.g. Figure 1 in \cite{Chenetal06}),
where we know that a positive downscale flux of energy is only transient and cannot be 
constant (e.g. Figure 1 in \cite{Boffetta07}). Such distinction between transient and constant fluxes 
is imperative to modeling efforts. In the former case, there is no cascade or enhancement of dissipation due to
turbulence whereas in the latter case, dissipation becomes independent of Reynolds number. 

This issue is especially important when drawing conclusions
from limited resolution simulations, true of most cases including ours. 
A constant flux indicates that dissipation should be independent of the simulation resolution.
One may contend that a constant flux is just a consequence of a steady state
and having forcing localized to the largest scales.
However, it may very well be that the flow
reaches steady state due to direct viscous dissipation acting on all scales
as shown in \cite{Herringetal94} for potential enstrophy rather than a 
Kolmogorov-like inertial cascade.

\begin{figure}[htpb]
\centering
\includegraphics[totalheight=0.6\textheight,width=0.7\textwidth]{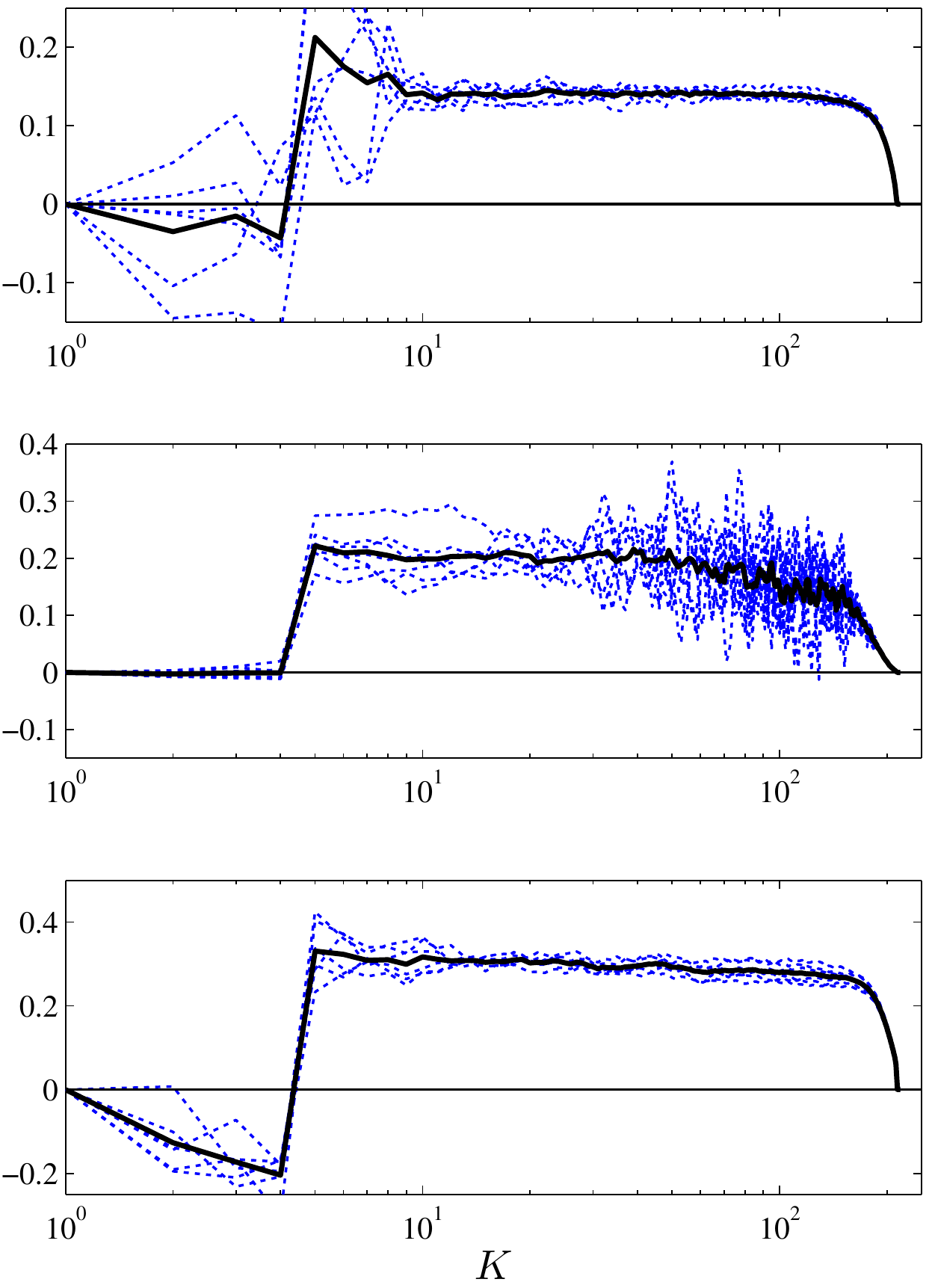}
\caption{Plots of total energy flux, 
$\langle \Pi_K \rangle$, from runs 
Rs (top), rS (middle), and RS (bottom).
Dotted (blue) lines are fluxes taken from an instantaneous time snapshot. Solid (black) lines
are averaged over time.}
\lb{Fig:Flux_All_iso}\end{figure}

We also compute the potential enstrophy flux. 
Figure \ref{Fig:Flux_All_PV2_iso}
shows that, in a manner similar to that of energy, there is a positive and constant flux of potential enstrophy 
in all three extreme cases\footnote{
The flux $\langle\Pi^Q_K\rangle$ in the rS case is noisy at small scales for the same reasons
as in Figure \ref{Fig:Flux_All_iso} for $\langle\Pi_K\rangle$.}. 
The plots in Figure \ref{Fig:Flux_All_PV2_iso} are the first measurements
of potential enstrophy flux in rotating stratified Boussinesq flows and constitute one of the main results
in this Letter. They can be regarded as the first empirical confirmation of analytical results in
\cite{Kurienetal06} which derived an exact law for potential enstrophy flux in physical space as a function of scale. 
Unlike for energy, $\langle\Pi^Q_K\rangle$ in Figure \ref{Fig:Flux_All_PV2_iso} for the RS case is not negative over $K\le 4$.

\begin{figure}[htpb]
\centering
\includegraphics[totalheight=0.6\textheight,width=0.7\textwidth]{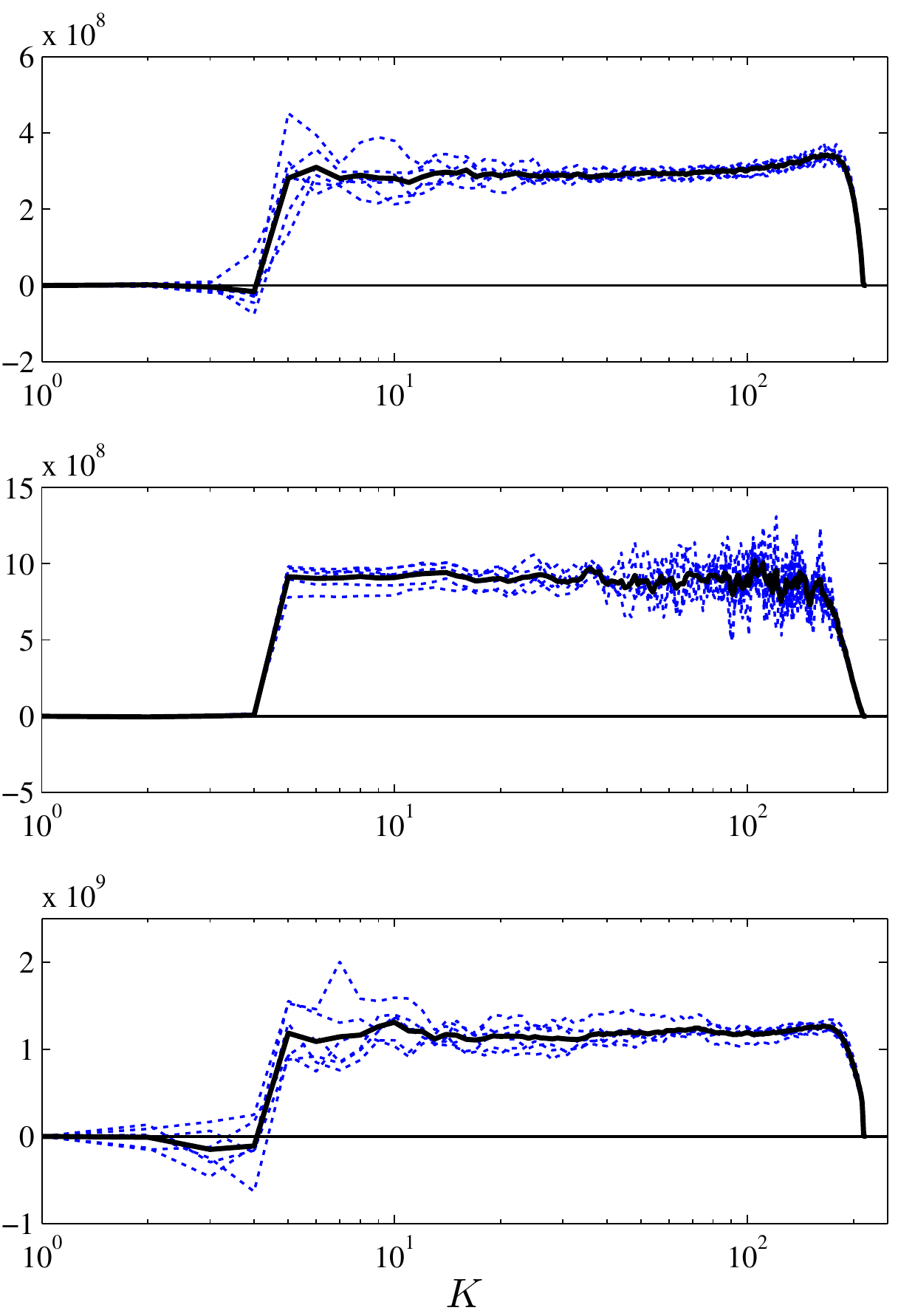}
\caption{Plots of potential enstrophy flux, $\langle \Pi^{Q}_K \rangle$, from runs 
Rs (top), rS (middle), and RS (bottom).
Dotted (blue) lines are fluxes taken from an instantaneous time snapshot. Solid (black) lines
are averaged over time.}
\lb{Fig:Flux_All_PV2_iso}\end{figure}

Rotating and stratified flows are often said to be `two-dimensionalized' in some sense, eliciting comparisons with two-dimensional turbulence and often justifying the study of the latter as a simplified paradigm for geophysical flows. Here we point out that the existence of a concurrent flux of both energy and potential enstrophy in rotating and stratified flow to smaller scales is in itself
a marked departure of these flows from 2D turbulence. In the latter case, it is known 
(e.g. \cite{Kraichnan67,Boffetta07}) that the two cascades cannot co-exist over the same scale-range since 
a forward cascade of enstrophy acts as a constraint leading to an inverse cascade 
of energy to larger scales. 

In rotating and stratified flows, the velocity and scalar fields
can be decomposed into the sum of a vortical component, which accounts for all the potential vorticity
in the flow, and into a wave component, which has zero potential vorticity (e.g. \cite{Bartello95,SmithWaleffe02}).
In the strongly rotating and strongly stratified regime, it is 
known (see \cite{Bartello95,Babinetal97,EmbidMajda98})
that the vortical component of the flow is governed by quasigeostrophic dynamics, which
is very similar to 2D turbulence \cite{Charney71}. In this regime, a forward energy cascade 
of the vortical component is suppressed due to the forward cascade of potential enstrophy.
We verified (to appear in \cite{AluieKurien_long}) 
that this is indeed the case in our RS run, where the forward energy cascade 
(in Figure \ref{Fig:Flux_All_iso}, bottom panel) is due to 
the wave component of the flow in agreement with  \cite{Bartello95}.
However, the situation in the remaining two cases, runs Rs and rS, is markedly different from both 2D and quasigeostrophic dynamics as will be discussed in \cite{AluieKurien_long}.

\section{Conclusions}
The two main results presented in this Letter are (i) energy and potential enstrophy budgets which resolve the
dynamics simultaneously in space and in scale and 
(ii) concurrent and persistent cascades of energy and potential energy to small scales in
three extreme cases of rotating and stratified Boussinesq flow simulations.
The numerical results on constant fluxes of potential enstrophy
constitute the first direct empirical evidence in support of analytical results 
by \cite{Kurienetal06}. Our findings show a clear departure of the flows
we study from 2-dimensional turbulence and should be incorporated in
any phenomenological treatments of strongly rotating and/or stratified Boussinesq 
flows. In a longer forthcoming work \cite{AluieKurien_long}, we shall refine our analysis
to study anisotropy of these cascades, their pointwise and scale-locality properties, and 
quantify contributions from vortical and wave components.

\vspace{0.4cm}
\noindent {\small
{\bf Acknowledgements.} 
This research used resources of the Argonne Leadership Computing Facility at Argonne
National Laboratory, which is supported by the Office of Science of
the US  DOE under contract DE- AC02-06CH11357.
HA acknowledges partial support from NSF grant PHY-0903872 during a visit to the 
Kavli Institute for Theoretical Physics.
This research was performed under the auspices of the US DOE 
at LANL under Contract No. DE-AC52-06NA25396. HA was supported
by the LANL/LDRD program and by the DOE ASCR program in Applied 
Mathematical Sciences. SK received partial funding from NSF program Collaborations in the Mathematical 
Geosciences: NSF CMG-1025188.
}

\section{Appendix: Budgets\label{ap:Budgets}}
\vspace{-0.5cm}
For the sake of completion, we write down the complete expression for the transport term in (\ref{largeE}):
\begin{eqnarray}
J_j(\bx)  &=&  \OL{u}_j \frac{|\OL\bu|^2}{2} + \OL{u}_i~\OL\tau(u_i,u_j) - \nu\partial_j\frac{{|\OL\bu|^2}}{2},\\
&+&  \OL{u}_j \frac{|\OL\theta|^2}{2} + \OL\theta~\OL\tau(\theta,u_j) - \kappa\partial_j\frac{{|\OL\theta|^2}}{2}.
\nonumber\end{eqnarray}

In large-scale potential enstrophy budget (\ref{largePV2}), $\bJ^{Q}(\bx)$ is space transport, 
$D^{Q}_\ell(\bx)$ is dissipation, and $\epsilon_{inj}^{Q}$ is the potential enstrophy injected due
to external forcing. These are defined as
\begin{eqnarray}
D^{Q}_\ell(\bx) &=&\hspace{-3.0mm} \nu N^2 |\grad \OL\omega_z|^2 \hspace{-1.0mm}+\hspace{-1.0mm}\kappa f^2 | \grad \partial_z\OL\theta|^2 \hspace{-1.0mm}-\hspace{-0.2mm}(\nu\hspace{-1.0mm}+\hspace{-1.0mm}\kappa) fN(\grad \partial_z\OL\theta)\bdot \grad \OL\omega_z\hspace{+7.0mm}  \lb{PV2dissip}\\
J^{Q}_j (\bx) &=&\hspace{-3.0mm}\OL{u}_j\frac{|\OL{q}|^2}{2} \hspace{-1.0mm}+ \OL{q}~\OL\tau(u_j,q)\hspace{-1.0mm}+  \nu N~\OL{q}~\partial_j\OL{\omega}_z -\kappa f~\OL{q}~\partial_j\partial_z\OL{\theta}, ~~ \lb{PV2transport}\\
\epsilon_{inj}^{Q} (\bx)
&=&\hspace{-2.0mm} -N ~\OL{q}~ \OL{{\mathcal F}^\omega}_z + f~\OL{q}~\partial_z\OL{{\mathcal F}^\theta}
 \lb{PV2inject}
\end{eqnarray}   
Note that while the first two dissipation terms in (\ref{PV2dissip}) are positive definite, 
the last term can be of either sign.

\bibliographystyle{unsrt}
\bibliography{turbulence,geophysical}

\end{document}